\documentclass{article}
\usepackage{spconf,amsmath,graphicx}

\usepackage{enumitem}
\usepackage{multirow}
\usepackage[table,xcdraw]{xcolor}
\usepackage{graphicx}
\usepackage[colorlinks]{hyperref}
\usepackage{amssymb}
\setlist{nosep, leftmargin=14pt}

\usepackage{mwe} 

\title{Duplex Contextual Relation Network for Polyp Segmentation}
\name{Zijin Yin, Kongming Liang, Zhanyu Ma, Jun Guo}
\address{Beijing University of Posts and Telecommunications, Beijing, China}
\begin{document}
%
\maketitle
\begin{abstract}
Automatic polyp segmentation from colonoscopy has a pivotal role in the early diagnosis and surgery of Colorectal Cancer(CRC). However, the diversity of polyps across different images significantly increases the difficulty of accurate polyp segmentation. Existing researches focus on learning the contextual information within an individual image but fail to exploit the co-occurrent visual patterns of polyps across images. In this paper, we argue that exploring contextual correlation from a holistic view of the whole dataset is essential and propose a Duplex Contextual Relation Network (DCRNet) to capture both within-image and cross-image contextual relations. Based on the above two types of similarity, the feature of each input region can be enhanced by its contextual region embedding within and across images. To store the characteristic region embedding from previous images during training, an episodic memory is designed and operates as a queue. We evaluate the proposed method on the EndoScene, Kvasir-SEG, and the recently released large-scale PICCOLO dataset. Experimental results show that our proposed DCRNet outperforms the state-of-the-art methods in terms of the widely-used evaluation metrics.
\end{abstract}


\section{Introduction}
\label{sec:intro}

\begin{figure}[t]
\includegraphics[width=\columnwidth]{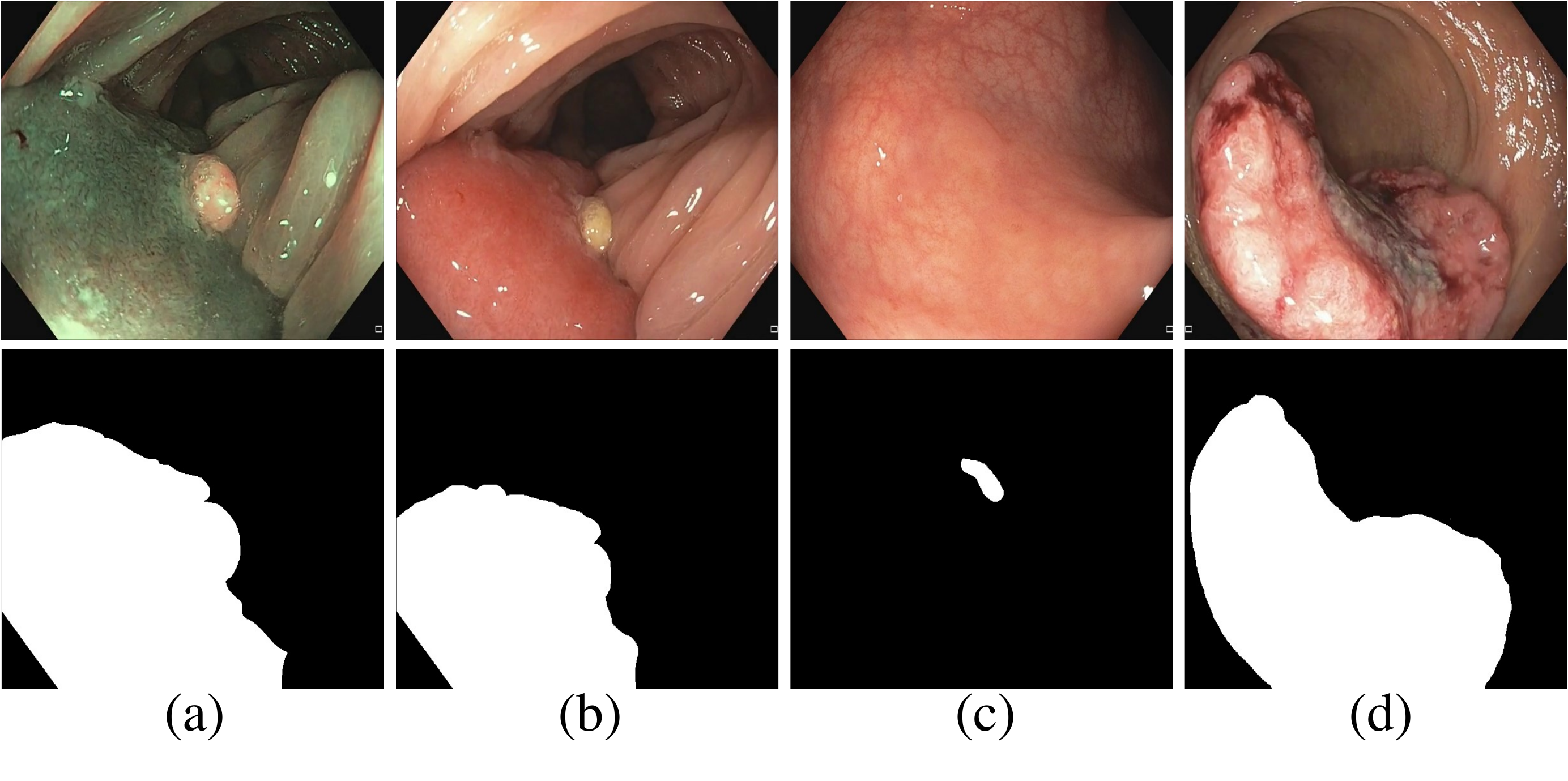}
\caption{Illustration of polyps in colonoscopy: (a)\&(b) color inconsistency, (b)\&(c) size diversity, (b)\&(d) texture diversity, and (c) low contrast to surroundings.}\label{intro}
\end{figure}

The automatic polyp segmentation technique plays an important role in addressing the issue of prevention of Colorectal Cancer (CRC). It can locate polyps from colonoscopy and significantly reduce manual labor and decline the misdiagnosis rate.
However, automatic polyp segmentation has always been a challenging task mainly for two reasons:
(i) polyps generally vary in appearances such as size, color, and texture; and (ii) the boundary between polyp and mucosa is usually blurred. Some examples are shown in Figure.\ref{intro}.

Some previous methods focus on extracting multi-scale features to address the above issues. For example, ACSNet \cite{ACSNet} combines the global context and local details to deal with the shape and size variance of polyps regions. PraNet \cite{PraNet} aggregates multi-scale features and successively refines the segmentation map by extracting silhouettes according to the local features. Another line of work explicitly leverages auxiliary information to constrain the segmentation results. For instance, SFANet \cite{fang2019selective} employs area-boundary constraints to improve the segmentation performance of both polyp regions and boundaries. However, previous efforts mainly focus on context information within an individual image and ignore the cross-data semantic similarity.

In clinical application, co-occurrent visual patterns widely exist across different images. For example, Figure.\ref{intro}(a)\&(b) illustrate that samples collected under different illumination conditions are inconsistent in colors but resembled in visual structures.
Meanwhile, \cite{mirasadi2019content} has proven the significance of retrieving from other images in the procedure of lesion treatment in radiology. And the superiority of cross-image modeling has been demonstrated in metric learning \cite{wang2020cross,oord2018representation} but is rarely discussed in the segmentation task.
Motivated by the above thoughts and observations, we propose to explore contextual relations from the holistic perspective of the whole dataset. The cross-image consistency of the same semantic class is exploited to delineate the co-occurrent visual patterns.

\begin{figure*}[t]
\includegraphics[width=\textwidth]{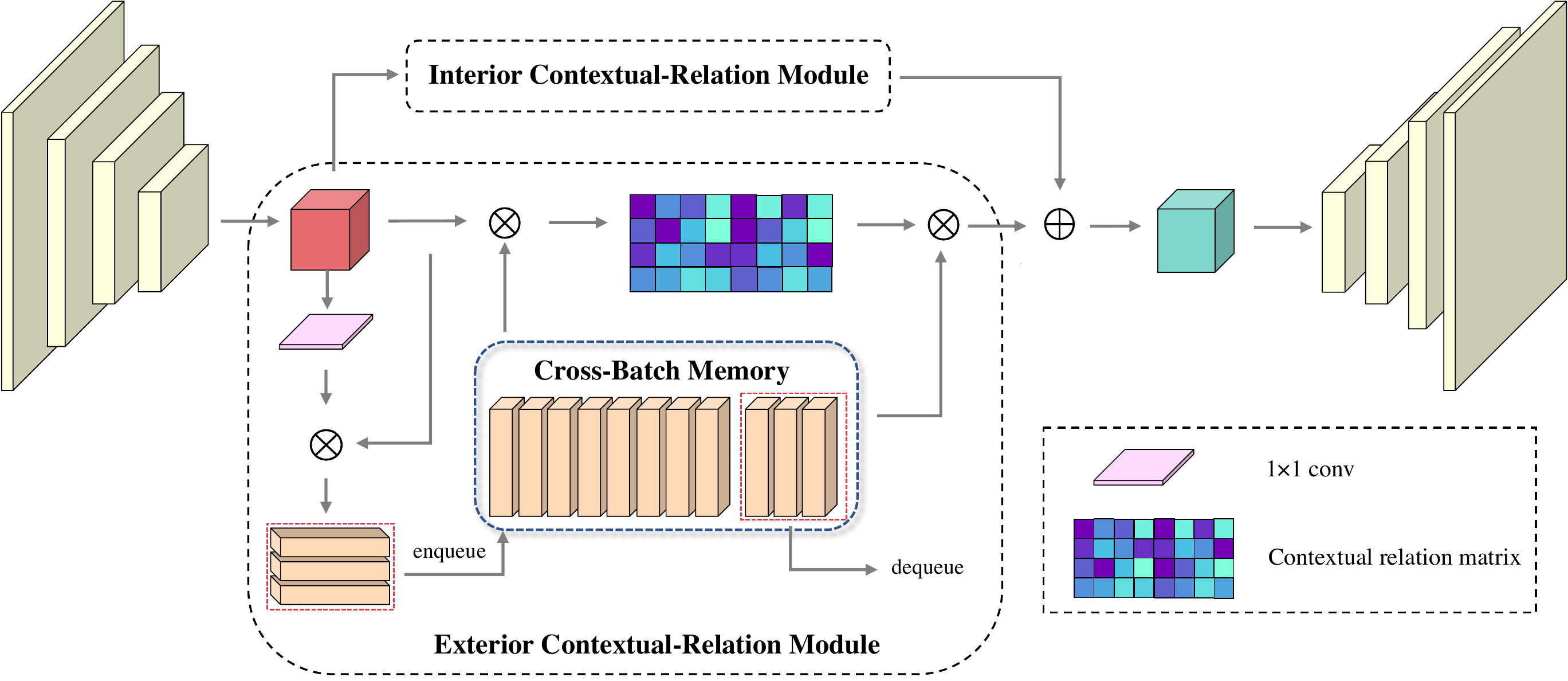}
\caption{An overview of the proposed DCRNet}\label{fig1}
\end{figure*}

In this paper, a Duplex Contextual Relation Network (DCRNet) is proposed to simultaneously capture the contextual relations across images and within individual images. Specifically, we design two parallel attention-based modules which can be incorporated into any encoder-decoder architecture (e.g U-Net\cite{ronneberger2015u}). The first module is called \emph{Interior Contextual-Relation Module} which estimates the similarity between each position and all the positions within the same image. The feature at one position is further aggregated by the features at all positions according to the estimated similarity. The second module is called \emph{Exterior Contextual-Relation Module} which estimates the similarity between each position and the positions across different images, thus meliorating intra-class consistency and inter-class separability. Based on the across-image similarity, the feature at one position can be further enhanced by the contextual region embedding from other images through feature aggregation. To achieve this, we employ a memory named \emph{Region Cross-Batch Memory} that operates as a queue to store the characteristic region embedding of previously seen images from past training epochs. Therefore, similar features can be related to each other even though they come from different samples.

In summary, our contributions mainly are: 1) We propose a novel scheme by capturing the contextual relations across images and within an individual to delineate the co-occurrent visual patterns. 2) A memory is designed and operates as a queue to store the characteristic region embedding from previous images. 3) Extensive experiments demonstrate that the proposed DCRNet outperforms the state-of-the-art methods on three public datasets.


\section{Method}
\subsection{Overview}
As illustrated in Figure.\ref{fig1}, the features from the last encoder backbone which is utilized as ResNet-34, are processed by two parallel modules: \emph{Interior Contextual-Relation Module (ICR)} and \emph{Exterior Contextual-Relation Module (ECR)}. The proposed memory called \emph{Region Cross-Batch Memory (ROM)} works as a queue to store up-to-date embedding features from previous mini-batches. Finally, we fuse the features from two blocks to obtain augmented representations for pixel-level prediction.


\subsection{Interior Contextual-Relation Module}
As demonstrated by Fu \emph{et al.}\cite{fu2019dual}, the self-spatial-attention mechanism could adaptively integrate local features with their global dependencies. Following this design, we adopt Position Attention Module $s(\cdot)$\cite{fu2019dual} to capture contextual relations between any two pixels. Specifically, given a feature $\mathbf{A} \in \mathbb{R}^{C \times H \times W}$ from the encoder, the augmented region representation is computed by $\mathbf{B} = s(\mathbf{A}) \in \mathbb{R}^{C \times H \times W}$.

After $s(\cdot)$\cite{fu2019dual}, the representation at each position is a weighted sum of that of all pixels. With such an adaptive mechanism, the pixel representations own stronger contextual relations with the same semantic class surroundings, and weaker with the different semantic class areas.

\subsection{Exterior Contextual-Relation Module}
In a clinical setting, there is the existence of co-occurrent visual patterns of polyps across different samples. Based on this critical observation, region features belonging to the same semantic class of all training data should have contextual relations. Thus, we propose a novel contextual relation exploring module across different samples.

\noindent \textbf{Contextual region embedding.} For a given feature $\mathbf{A} \in \mathbb{R}^{C \times H \times W}$, we first utilize a transformation function $\psi(\cdot)$, which is implemented by 1 × 1 conv → BN → ReLU, to compute a coarse segmentation map $\mathbf{M} \in \mathbb{R}^{1 \times H \times W}$, where each entry indicates the degree that the corresponding pixel belongs to the polyp region. Then the contextual region embedding is computed as below:
\begin{equation}\label{regionEmbedding}
   \mathbf{E} = \phi(\mathbf{M})^{\top} \cdot \phi(\mathbf{A})
\end{equation}
Here, $\phi(\cdot)$ is flatten function, and $\cdot$ is matrix multiplication.
$\mathbf{E} \in \mathbb{R}^C$. 

\noindent \textbf{Contextual relation matrix.} Suppose that the buffered region embeddings are 
$\mathbb{E} = \left\{\mathbf{E}_{1}, \mathbf{E}_{2}, \ldots, \mathbf{E}_{S}\right\} \in \mathbb{R}^{S \times C}$ 
where $S$ is the bank size, and flattened feature representations of current mini-batch are 
$\mathbb{A} = \left\{\phi(\mathbf{A}_{1}), \phi(\mathbf{A}_{2}), \ldots, \phi(\mathbf{A}_{B})\right\}$ where $B$ is the batchsize. 
Then we perform a matrix multiplication between them and apply a softmax layer to calculate the contextual attention map $\mathbb{X} \in \mathbb{R}^{HW \times B \times S}$:
\begin{equation}\label{attentiom map}
    x_{z j i}=\frac{\exp \left(\mathbb{E}_{i} \cdot \mathbb{A}_{j z}\right)}{\sum_{i=1}^{N} \exp \left(\mathbb{E}_{i} \cdot \mathbb{A}_{j z}\right)}
\end{equation}
where $x_{z j i}$ measures contextual relation in the $z^{th}$ pixel between $i^{th}$ image and $j^{th}$ image. Note that the more similar feature representations of the two images contributes to greater correlation between them.

\noindent \textbf{Augmented representations.}
The final augmented feature representations is computed by:
\begin{equation}\label{augment representation}
    \mathbb{Y} = \rho(\delta(\mathbb{X} \cdot \mathbb{E})) \in \mathbb{R}^{B \times C \times H \times W} 
\end{equation}
where $\delta(\cdot)$ is the transpose function used to adjust the dimension order, and $\rho(\cdot)$ is the unflatten function used to recover spatial dimension.

\subsection{Region Cross-Batch Memory}
Inspired by non-parametric memory modules for embedding learning and contrastive learning \cite{wang2020cross,he2020momentum}, since we probe into the mutual contextual relations between different region embeddings across mini-batches, a memory concept is adopted and hence used to store previously seen embeddings.  Furthermore, the work in \cite{wang2020cross} revealed “slow drift” phenomena which signify features drift exceptionally slow even as the model parameters are updating throughout the training process. The above discovery indicates the past mini-batches can be a considerably important resource, especially in medical computation. However, the embeddings too far away from the current mini-batch could cause feature-level inconsistency, which implies that the past entities should be iteratively discarded. Therefore, we operate the memory bank as a queue with a first-in-first-out principle.

Specifically, at the early stage of training, we initialize the memory by ﬁlling all the calculated contextual region embeddings. When the number of elements reaches the bank size $S$, we enqueue the region embeddings of the current mini-batch and dequeue the entities of the earliest mini-batch. Signiﬁcantly, the setting of band size $S$ should be moderate: excessive small size could not arise rich contextual relations and excessive large size usually cause out-of-date data. The Region Cross-Batch Memory will be removed as inference.

\subsection{Loss Functions}
Similar to \cite{PraNet,ACSNet}, we adopt the deep supervision strategy for three intermediate maps of decoder branch and coarse segmentation map $M$. And each is up-sampled to the same size as the label. For the loss function, we employ the combination of a weighted binary cross-entropy (BCE) loss $\mathcal{L}_{w b c e}$ \cite{wei2020f3net} and a Dice loss $\mathcal{L}_{D i c e}$. 
Using this strategy could help the model consider both pixel-level and region-level measurements.

\section{Experiments}
\begin{table}[t]
\caption{Quantitative results on three benchmarks.}\label{results}
\centering
\setlength{\tabcolsep}{1.3mm}{
\begin{tabular}{lr|r|ccccc}
\hline
 & Methods & \multicolumn{1}{c}{MAE} & \multicolumn{1}{c}{Dice} & \multicolumn{1}{c}{IoU} & \multicolumn{1}{c}{$\mathcal{F}$} & \multicolumn{1}{c}{$S_{\alpha}$} \\ \hline
 & U-Net\cite{ronneberger2015u}  & \multicolumn{1}{c}{4.4}  & 73.78  & 66.54 & 68.78 & 83.54   \\
 & U-Net++\cite{zhou2018unet++}  & \multicolumn{1}{c}{4.5}  & 72.88  & 64.58  & 63.68  & 82.41\\
 & ResUNet++\cite{jha2019resunet++} & \multicolumn{1}{c}{6.3}  & 52.41  & 44.33  & 43.60  & 71.02\\
 & PraNet\cite{PraNet} & \multicolumn{1}{c}{3.5} & 81.73  & 74.38   & 75.79  & 88.00\\
 & ACSNet\cite{ACSNet}   & \multicolumn{1}{c}{3.0}  & 85.15 & 78.67 & 81.58   & 90.54   \\
\multirow{-6}{*}{\rotatebox{90}{EndoScene}} & \textbf{Ours} & \multicolumn{1}{c}{\textbf{3.0}} & \textbf{85.41}  & \textbf{78.86} & \textbf{83.20}  & \textbf{90.79}  \\ \hline
\end{tabular}
}

\centering
\setlength{\tabcolsep}{1.3mm}{
\begin{tabular}{lr|r|ccccc}
\hline
 & Methods & \multicolumn{1}{c}{MAE} & \multicolumn{1}{c}{Dice} & \multicolumn{1}{c}{IoU} & \multicolumn{1}{c}{$\mathcal{F}$} & \multicolumn{1}{c}{$S_{\alpha}$} \\ \hline
 & U-Net\cite{ronneberger2015u} & \multicolumn{1}{c}{4.2} & 85.97   & 78.70   & 73.13   & 88.36  \\
 & U-Net++\cite{zhou2018unet++}& \multicolumn{1}{c}{5.2} & 84.16 & 76.02 & 70.33  & 87.17 \\
 & ResUNet++\cite{jha2019resunet++}& \multicolumn{1}{c}{5.6} & 81.09 & 72.73 & 64.75    & 85.22 \\
 & PraNet\cite{PraNet} & \multicolumn{1}{c}{3.1} & 89.20   & 83.61  & 77.97  & 90.96 \\
 & ACSNet\cite{ACSNet} & \multicolumn{1}{c}{3.2} & 89.32 & 83.83   & 79.04  & 90.96  \\
\multirow{-6}{*}{\rotatebox{90}{Kvasir-SEG}} & \textbf{Ours} & \multicolumn{1}{c}{\textbf{2.9}} & \textbf{90.14}   & \textbf{84.44}   & \textbf{82.05} & \textbf{91.49} \\ \hline
\end{tabular}
}

\centering
\setlength{\tabcolsep}{1.3mm}{
\begin{tabular}{lr|r|ccccc}
\hline
 & Methods & \multicolumn{1}{c}{MAE} & \multicolumn{1}{c}{Dice} & \multicolumn{1}{c}{IoU} & \multicolumn{1}{c}{$\mathcal{F}$} & \multicolumn{1}{c}{$S_{\alpha}$} \\ \hline
 & U-Net\cite{ronneberger2015u} & \multicolumn{1}{c}{5.0}  & 66.81  & 60.59   & 57.04   & 79.12 \\
 & U-Net++\cite{zhou2018unet++} & \multicolumn{1}{c}{5.4}  & 68.21  & 61.48  & 58.11 & 79.07 \\
 & ResUNet++\cite{jha2019resunet++}  & \multicolumn{1}{c}{5.8} & 60.24  & 53.68  & 47.04  & 75.15 \\
 & PraNet\cite{PraNet} & \multicolumn{1}{c}{3.0} & 75.34 & 69.77  & 65.88  & 84.71 \\
 & ACSNet\cite{ACSNet} & \multicolumn{1}{c}{2.6}  &83.49  &77.88 &75.04  &-  \\
\multirow{-6}{*}{\rotatebox{90}{PICCOLO}} & \textbf{Ours} & \multicolumn{1}{c}{\textbf{2.0}} & \textbf{85.13}  &\textbf{79.43} & \textbf{78.09} & \textbf{89.70} \\ 
\hline
\end{tabular}
}
\end{table}
\subsection{Datasets}
Experiments are conducted on three polyp segmentation datasets: EndoScene \cite{EndoScene}, Kvasir-SEG \cite{Kvasir-SEG}, and PICCOLO \cite{PICCOLO}. EndoScene contains 912 manually segmented White-Light images. We use the default split for training, validation, and testing. Kvasir-SEG contains 1000 White-Light images with pixel-level manual labels. We randomly choose 60\% of the dataset as the training set, 20\% as the validation set, and the remaining as the test set. The last is the recently released PICCOLO dataset, which contains 3433 manually annotated images (2131 White-Light images and 1302 Narrow-Band images). 
We use the default data splitting, which is 2203 images for the training set, 897 images for the validation set, and 333 images for the test set.
All the images are resized to 224 × 224 in our experiments.

\subsection{Implementation Details and Evaluation Metrics}
Our model is implemented in Pytorch and trained on a single NVIDIA RTX 2080Ti. We employ the Adam optimizer with a learning rate of 1e-4 for 150 epochs. And the batch size is set as 4 for all datasets. We also utilize data augmentation strategies such as vertical and horizontal random flips, zoom, shift, and rotation. The memory size is set to 20 for Kvasir-SEG, and 40 for EndoScene and PICCOLO. 

Following \cite{ACSNet, PraNet}, we use three basic evaluation metrics including ``MAE'', ``Dice'', ``IoU''. 
And we further introduce a metric boundary F-measure $\mathcal{F}$\cite{boundary-F} to measure the contour accuracy and adopt $S_{\alpha}$\cite{fan2017structure} to measure the global structural similarity between prediction and Ground-Truth.

\begin{table}[h]
\caption{Model and inference analysis on PICCOLO.}\label{Inference}
\centering
\begin{tabular}{r|c|c|c}
\hline
Methods     & Inference(FPS) & Model size(MB) & Dice \\ \hline
PraNet      & $\sim$24fps    & 30.5MB         & 75.34     \\
ACSNet      & $\sim$22fps    & 29.5MB         & 83.49     \\
\textbf{Ours} & \textbf{$\sim$53fps}    & \textbf{28.7MB}        & \textbf{85.13}     \\ \hline
\end{tabular}
\end{table}

\begin{figure}[t]
    \centering
    \includegraphics[width=\columnwidth]{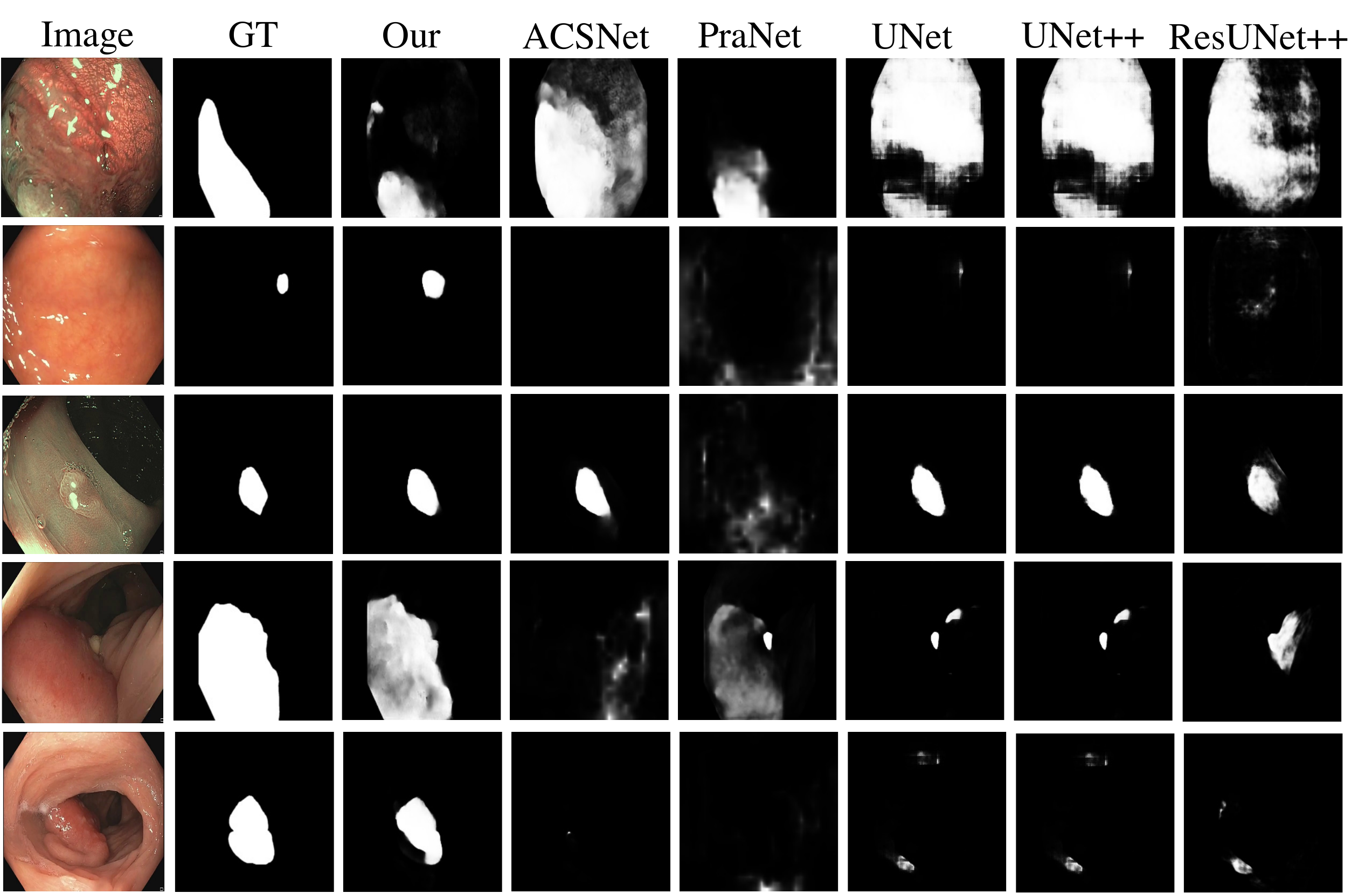}
    \caption{Qualitative comparison between different methods on PICCOLO, showing our method can better handle diverse scenarios, such as blurred boundary, brightness change, enormous and tiny polyp.}
    \label{qualitative-result}
\end{figure}

\begin{table}[t]
\caption{Ablation Study for DCRNet on EndoScene and PICCOLO datasets.}\label{ablation}
\centering
\begin{tabular}{ccc|cc|cc}
\hline
\multicolumn{1}{c}{\multirow{2}{*}{ICR}} & \multicolumn{1}{c}{\multirow{2}{*}{ECR}} & \multicolumn{1}{c|}{\multirow{2}{*}{ROM}} & \multicolumn{2}{c|}{EndoScene} & \multicolumn{2}{c}{PICCOLO} \\
\multicolumn{1}{c}{} & \multicolumn{1}{c}{} & \multicolumn{1}{c|}{} & Dice & IoU & Dice & IoU \\ \hline
\checkmark &  &  & 85.23 & 78.88 & 79.27 & 73.38 \\
 & \checkmark &  & 83.61 & 76.69 & 79.7 & 74.12 \\
 & \checkmark & \checkmark & 84.93 & 78.31 & 82.01 & 76.22 \\
\checkmark & \checkmark & \checkmark & 85.41 & 78.86 & 85.12 & 79.31 \\ \hline
\end{tabular}
\end{table}

\subsection{Results and Analysis}
\noindent \textbf{Comparison to the State-of-the-Art Methods.}
We compare our DCRNet with three medical image segmentation methods: U-Net\cite{ronneberger2015u}, U-Net++\cite{zhou2018unet++}, ResUNet++\cite{jha2019resunet++}, and two SOTA polyp segmentation methods: PraNet\cite{PraNet} and ACSNet\cite{ACSNet} whose results are reproduced using the official released code with default settings. As shown in Table.~\ref{results}, obviously, our model achieves superior performance in terms of all metrics on three benchmarks. And the considerable margins over \emph{IoU}, \emph{Dice}, and \emph{$\mathcal{F}$} suggest that our method is significantly more accurate for both region and boundary. Since the PICCOLO\cite{PICCOLO} has the most data and the most complex scene in all publicly available datasets, the excellent performance on it could better report the robustness of our method in realistic clinical application.

\noindent \textbf{Inference and model analysis.}
In Table.~\ref{Inference}, we evaluate the inference time and model parameters of DCRNet and other SOTA algorithms with the same batch size of 4 and the 1080Ti GPU platform. As shown, our method runs drastically faster than others and owns the minimum number of parameters. This verifies that our model is more appropriate for clinical applications.

\subsection{Ablation Study}
To validate the effectiveness and necessity of each module in our proposed method, we compare DCRNet with its three variants in Table.~\ref{ablation}. Specifically, the Backbone refers to the original U-Net with pretrained ResNet-34 encoder, and we successively add ICR, ECR, and ROM to it.
As shown, with the progressive introduction of each component, our algorithm has witnessed a certain degree of performance improvement, boosting Dice by 0.68\%, 1.11\%, 2.31\% respectively. It is noteworthy that the improvement brought by Backbone + ECR + ROM on PICCOLO is more remarkable than that on EndoScene. This observation indicates that simulating such cross-image clinical diagnosis is momentous, and confirms the effectiveness and importance of our core thought.

\section{Conclusion}
In this paper, we propose Duplex Contextual Relation Network (DCRNet) to explore co-occurrent visual patterns of polyps across images. The proposed network contains two parallel modules which are utilized to capture contextual relations within and cross images respectively. To store the characteristic region embedding from previous images, a memory is designed to operate as a queue. Experimental results show that the proposed method achieves state-of-the-art performance on three datasets in terms of the widely-used evaluation metrics. Ablation studies are conducted to demonstrate the effectiveness of each proposed component.

\bibliographystyle{IEEEbib}
\bibliography{strings,refs}

\end{document}